\documentclass[aps,twocolumn,a4paper,showpacs,superscriptaddress]{revtex4}
\usepackage{graphicx}
\usepackage[all]{xy}
\usepackage{amsmath}
\usepackage{amssymb}
\newcommand{\be}{\begin{equation}}
\newcommand{\ee}{\end{equation}}
\newcommand{\ben}{\begin{eqnarray}}
\newcommand{\een}{\end{eqnarray}}
\newcommand{\bes}{\begin{subequations}}
\newcommand{\ees}{\end{subequations}}

\newcommand{\bb}{\bibitem}
\newcommand{\sech}{{\rm sech}}
\newcommand{\arcsinh}{{\rm arcsinh}}

\newcommand{\nn}{\nonumber\\}

\begin{document}
\title{Braneworld solutions for modified theories of gravity with non-constant curvature}
\author{D. Bazeia} 
\affiliation{Departamento de F\'\i sica, Universidade Federal da Para\'\i ba, 58051-970 Jo\~ao Pessoa, PB, Brazil}
\author{ A. S. Lob\~ao Jr}
\affiliation{Escola T\' ecnica de Sa\'ude de Cajazeiras, Universidade Federal de Campina Grande, 58900-000 Cajazeiras, PB, Brazil}
\author{L. Losano}
\affiliation{Departamento de F\'\i sica, Universidade Federal da Para\'\i ba, 58051-970 Jo\~ao Pessoa, PB, Brazil}
\author{R. Menezes}
\affiliation{Departamento de Ci\^encias Exatas, Universidade Federal da Para\'\i ba, 58297-000 Rio Tinto, PB, Brazil}
\affiliation{Departamento de F\'\i sica, Universidade Federal de Campina Grande, 58109-970 Campina Grande, PB, Brazil}
\author{Gonzalo J. Olmo}
\affiliation{Departamento de F\'\i sica, Universidade Federal da Para\'\i ba, 58051-970 Jo\~ao Pessoa, PB, Brazil}
\affiliation{Departamento de F\'\i sica Te\'orica and IFIC, Centro Mixto Universidad de Valencia - CSIC. Universidad de Valencia, Burjassot-46100, Valencia, Spain}

\begin{abstract}
We study braneworld models in the presence of scalar field in a five-dimensional geometry with a single extra dimension of infinite extent, with gravity modified to include a function of the Ricci scalar. We develop a procedure that allows to obtain analytical solution for the braneworld configuration in a diversity of models, in the much harder case where the Ricci scalar is non constant quantity.  
\end{abstract}


\maketitle

\section{Introduction}

The idea that our four-dimensional Universe may be viewed as a subspace in a higher-dimensional scenario has attracted the attention of the scientific community working in different fields, from elementary particle physics to cosmology, over the last fifteen years. In the first consistent models, space-time was described as a five-dimensional anti de Sitter ($AdS_5$) geometry, with a single extra spatial dimension of infinite extent \cite{rs}. This is known as the thin braneworld scenario. This setting was soon modified to give rise to a braneworld configuration sourced by a scalar field, giving rise to what is generally known as the thick brane scenario \cite{gw,f,c}. The thick brane scenario has been explored from many different perspectives \cite{g,bc,ca,b}. In particular, given our ignorance about the gravitational dynamics in higher-dimensional backgrounds, models of the $F(R)$ type, where $R$ is the Ricci scalar, have also been studied in several works, e.g., \cite{rev1,rev2,rev3,Odi1,Odi2,bR,Dz,Zhong,bd,aq,bRR,liu}. Some of these studies allowed us to test the robustness of the thick braneworld scenario under modifications of the gravitational dynamics away from the standard Hilbert-Einstein action. Exploration of linear stability against tensorial perturbations is then possible if we have control on the background solutions. It is for this reason that finding exact solutions of different configurations becomes a useful tool to test the robustness of the thick braneworld scenario. 

In this work, we follow the lines of \cite{bRR} and consider braneworld models described by theories of gravity of the $F(R)$ type, but we focus on the general case of non-constant scalar curvature $R$. For the sake of generality, we take the scalar field with standard and generalized kinematics, but we start with the simpler possibility, with scalar field models having standard kinematics. Despite the freedom allowed by this approach, we show that it is possible to find analytical solutions for several specific models using a well defined procedure which consists on requiring that the warp factor be proportional to a hyperbolic function of the extra dimension. The possibility to obtain analytical solutions with this approach allows us to study the stability of the gravity sector. In particular,  we find that the models considered are robust against fluctuations in the metric.  

To make the investigation easier to understand, we organize the subject of the work as follows. In Sec.~II we investigate the braneworld model with a polynomial modification of the Einstein-Hilbert action on general grounds, with the source scalar field having standard kinematics. We then study several specific models to illustrate that the procedure to generate analytical solutions works nicely. In Sec.~III we investigate the stability of the models worked out in Sec.~II, and verify their robustness against fluctuations in the metric sector. In Sec.~IV we extend our methodology to models in which the scalar field has generalized kinematics. This is much harder, but we show that the procedure still works, leading to the construction of exact solutions for several specific models. We end the work in Sec.~V with some comments and conclusions.

\section{braneworld model}

We start with a $5D$ model, with $F(R)$ gravity coupled to the real scalar field $\phi$ in the form
\be\label{Action}
{\cal S}=\int d^5x \sqrt{|g|}\left(-\frac14 F(R) + \frac12 \nabla_a \phi \nabla^a \phi - V(\phi)\right)\,.
\ee
We take $4\pi G^{(5)}=1$ and use $g=det(g_{ab})$. We investigate the case of a flat brane, with the line element given by $ds^2=e^{2A}\eta_{\mu\nu}-dy^2$ where $e^{2A}$ is the warp factor, $\eta_{\mu\nu}$ is the 4-dimensional Minkowski metric, and $y=x^4$ is the extra dimension. 

We suppose that both $A$ and $\phi$ are static and depend only on the extra dimension, such that $A=A(y)$ and $\phi=\phi(y)$. In this case, the variation of the action \eqref{Action} over the scalar field gives the following equation
\be\label{phi_equation}
\phi^{\prime\prime}+4A^\prime \phi^\prime =V_\phi\,,
\ee
where prime denotes derivative with respect to the extra dimension, and $V_\phi=dV/d\phi$. In this case we obtain the scalar curvature as
\begin{equation}
R=20A^{\prime2}+8A^{\prime \prime}\,.
\end{equation}
The equations obtained by the variation over the metric $g_{\mu\nu}$ can be combined to obtain the following equations 
\bes\label{E_equations}
\ben
\phi^{\prime2}\!\!\!&=&\!\!\!-\frac32 A^{\prime\prime}\! F_R \!+\! 4 \left(5A^{\prime2}\!A^{\prime\prime}\!\!-\!5A^{\prime\prime2}\!-\!4A^{\prime}\!A^{\prime\prime\prime}\!\!-\!A^{\prime\prime\prime\prime}\right) \!F_{RR}\!-\nn
\!\!\!&&\!\!\!-32 \left (5A^\prime A^{\prime\prime} + A^{\prime\prime\prime}\right)^2 F_{RRR}\,,\label{phiprime} \\
V(\phi) \!\!\!&=&\!\!\!  -\frac14 F(R) + \frac{1}{4} \left(5A^{\prime\prime}+8 A^{\prime2}\right) F_R -\nn
\!\!\!&&\!\!\!-2\left(35 A^{\prime2}A^{\prime\prime}+12A^\prime A^{\prime\prime\prime}+5A^{\prime\prime2}+A^{\prime\prime\prime\prime}\right)F_{RR} -\nonumber \\
\!\!\!&&\!\!\! -16  \left(5A^{\prime}A^{\prime\prime}+A^{\prime\prime\prime}\right)^2 F_{RRR}\,.
\label{potential}
\een
\ees 
Also, the energy density of the scalar field can be written in the general form 
\ben
\rho(y)= e^{2A}\!\!\!&\Big[&\!\!\!-\frac14 F(R) +\frac12 \left( A^{\prime\prime}+4A^{\prime2}\right)F_R- \nonumber\\
\!\!\!&&\!\!\!-4\left(15A^{\prime2}A^{\prime\prime}\!\!+\!5A^{\prime\prime2}\!\!+\!8A^\prime A^{\prime\prime\prime}\!\!+\!A^{\prime\prime\prime\prime}\right) \!F_{RR}\!-\nonumber\\ 
\!\!\!&&\!\!\!-32 \left(5A^\prime A^{\prime\prime}+A^{\prime\prime\prime}\right)^2 F_{RRR}\Big]\,.\label{ener_densi}
\een

For concreteness, in this work we choose to work with the family of gravitational Lagrangians 
\begin{equation}\label{eqFR}
F(R)=R+\alpha R^n\,,
\end{equation}
where $\alpha$ is a real number and $n$ is a positive integer. With this choice, the equations \eqref{E_equations} become
\bes\label{eqphiPot}
\ben
\phi^{\prime2}\!\!\!&=&\!\!\!-\frac32 A^{\prime\prime}-\frac{3n \alpha}2 A^{\prime\prime}\left(20A^{\prime2}+8A^{\prime \prime}\right)^{n-1} +\nn
\!\!\!&&\!\!\!+4\alpha n(n\!-\!1)\!\left(5A^{\prime2}\!A^{\prime\prime}\!\!-\!5A^{\prime\prime2}\!-\!4A^{\prime}\!A^{\prime\prime\prime}\!\!-\!A^{\prime\prime\prime\prime}\right) \times \nn
\!\!\!&&\!\!\!\times \left(20A^{\prime2}+8A^{\prime \prime}\right)^{n-2} \!-32\alpha n(n-1)(n-2) \times\nn
\!\!\!&&\!\!\! \times\left (5A^\prime A^{\prime\prime} + A^{\prime\prime\prime}\right)^2\left(20A^{\prime2}+8A^{\prime \prime}\right)^{n-3}\,,\label{eqphi}\\
V(\phi) \!\!\!&=&\!\!\!-3A^{\prime2}\!-\!\frac34A^{\prime \prime}\!-\!\frac14\alpha\left(20A^{\prime2}\!+\!8A^{\prime \prime}\right)^{n}  -\nn
\!\!\!&&\!\!\!+ \frac{\alpha n}{4} \left(5A^{\prime\prime}+8 A^{\prime2}\right) \left(20A^{\prime2}+8A^{\prime \prime}\right)^{n-1}-\nn
\!\!\!&&\!\!\!-2\alpha n(n\!-\!1)\left(35 A^{\prime2}A^{\prime\prime}\!+\!12A^\prime A^{\prime\prime\prime}\!+\!5A^{\prime\prime2}\!+\!A^{\prime\prime\prime\prime}\right)\times\nn
\!\!\!&&\!\!\!\times\left(20A^{\prime2}\!+\!8A^{\prime \prime}\right)^{n-2} \!-\!16\alpha n(n\!-\!1)(n\!-\!2)\times \nn
\!\!\!&&\!\!\!\times\left(5A^{\prime}A^{\prime\prime}\!+\!A^{\prime\prime\prime}\right)^2\left(20A^{\prime2}+8A^{\prime \prime}\right)^{n-3} \,.
\een
\ees 
The energy density then turns into 
\ben
\rho(y)\!\!\!&=&\!\!\!e^{2A}\Big[-3A^{\prime2}-\frac32A^{\prime \prime}-\frac14\alpha\left(20A^{\prime2}+8A^{\prime \prime}\right)^{n} +\nn
\!\!\!&&\!\!\!+\frac{\alpha n}2 \left( A^{\prime\prime}+4A^{\prime2}\right)\left(20A^{\prime2}+8A^{\prime \prime}\right)^{n-1}\!-\nn
\!\!\!&&\!\!\!-4\alpha n(n-1)\left(15A^{\prime2}A^{\prime\prime}\!\!+\!5A^{\prime\prime2}\!\!+\!8A^\prime A^{\prime\prime\prime}\!\!+\!A^{\prime\prime\prime\prime}\right) \times \!\nonumber\\ 
\!\!\!&&\!\!\!\times\left(20A^{\prime2}+8A^{\prime \prime}\right)^{n-2}-32\alpha n(n-1)(n-2)\times \nn
\!\!\!&&\!\!\!\times\left(5A^\prime A^{\prime\prime}+A^{\prime\prime\prime}\right)^2 \left(20A^{\prime2}+8A^{\prime \prime}\right)^{n-3} \Big]\,.\label{rhogeral}
\een

The challenge now is to find a consistent solution of the gravitational and scalar-field equations given above. To proceed, we assume that the space-time behaves asymptotically (when $y\to\infty$) as an anti-de Sitter (AdS) geometry. This requirement can be implemented by choosing the warp function in the form 
\begin{equation}\label{warpfact}
A(y)=B \ln \left(\sech(ky)\right)\,,
\end{equation}
where $B$ and $k$ are real and positive parameters. The positivity of the parameters is necessary for the graviton to be localized on the brane in the thin brane limit \cite{f,g}. We will next show that the ansatz (\ref{warpfact}) is good enough to allow us to obtain analytical solutions in several examples of current interest.

\subsection{First Example}\label{sec1}

\begin{figure}[t]
\begin{center}
\includegraphics[scale=0.72]{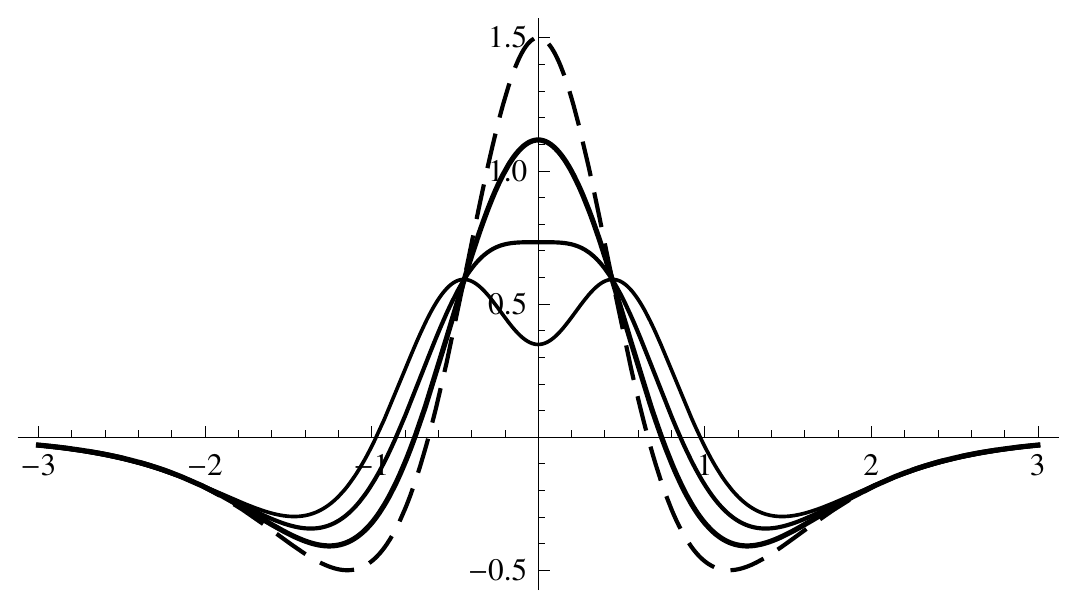}
\end{center}
\vspace{-0.5cm}
\caption{\small{Energy density \eqref{enerdensN2} depicted for $k=1$, $B=1$ and $\alpha=0$ (dashed line) , $\alpha=\alpha_s/2$ (thicker line), $\alpha=\alpha_s$ (thick line) and $\alpha=3\alpha_s/2$ (thinner line), with $\alpha_s=3/250$.\label{fig1}}}
\end{figure}

As a first example we consider a quadratic term in equation \eqref{eqFR} as done in \cite{bRR}. We take $F(R)=R+\alpha R^2$, and write the equations \eqref{eqphiPot} as 
\bes\label{EEM1}
\ben\label{phiprime_M1}
\!\phi^{\prime2}\!\!\!&=&\!\!\!-\frac32 A^{\prime\prime} 
\!-\!4\alpha \!\left(16A^{\prime\prime2}\!\!+\!5A^{\prime2}\!A^{\prime\prime}\!\!+\!8A^{\prime}\! A^{\prime\prime\prime}\!\!+\!2A^{\prime\prime\prime\prime}\right)\!,\\
V(\phi)\!\!\!&=&\!\!\!-\frac34 A^{\prime\prime}\!-\!3A^{\prime2}\!\\
\!\!\!&&\!\!\!-2\alpha \!\left(8A^{\prime\prime2}\!\!+\!69A^{\prime2}\!A^{\prime\prime}\!\!+\!10A^{\prime4}\!\!+\! 24A^{\prime} \!A^{\prime\prime\prime}\!\!+\!2A^{\prime\prime\prime\prime}\right)\!.~~~~~\nonumber\label{poten_M1}
\een
\ees
Also, the energy density \eqref{rhogeral} is simplified to the form
\ben\label{enerdensiM1}
\rho(y)\!\!\!&=&\!\!\! -e^{2A}\Big[3A^{\prime2}\!+\!\frac32A^{\prime\prime}\!+\!\\
\!\!\!&&\!\!\!+4\alpha \left(5A^{\prime4}\!+\!37A^{\prime2}\!A^{\prime\prime}\!+\!12A^{\prime\prime2}\!+\!16A^{\prime }\!A^{\prime\prime\prime}\!+\!2A^{\prime\prime\prime\prime}\right)\!\Big]\,.\nonumber
\een
Since the scalar field is real, the parameters $k$ and $B$ characterizing the warp factor and the scale $\alpha$ in the Lagrangian must satisfy the following constraint
\be\label{range}
\frac{-3}{8k^2(8+16B+5B^2)}\leq \alpha\leq \frac{3}{32k^2(1+4B)}.
\ee
We can now substitute equation \eqref{warpfact} in \eqref{phiprime_M1} to obtain the differential equation for the field $\phi$ as
\ben\label{eq1}
\!\phi^{\prime2}\!\!&=&\!\!a_1 \sech^2(k y)-a_2 \sech^4(k y)\,,
\een
where $a_1$ and $a_2$ are constants, given by
\bes\label{a12}
\ben
a_1\!\!&=&\!\!\frac32B k^2+4\alpha Bk^4 (8+16 B+5 B^2)\,, \\
a_2\!\!&=&\!\!4\alpha  Bk^4 (6 + B) (2 + 5 B)\,.
\een
\ees
The issue now is to obtain analytical solution of the equation \eqref{eq1} with a suitable choice of the parameters. For example, if we choose $\alpha=-3/[8k^2 (8+16 B+5 B^2)]$ we get from \eqref{eq1} the new equation
\begin{equation}
\phi^\prime=\pm \frac{k}{2} \sqrt{\frac{6B(6+B)(2+5B)}{8+16B+5B^2}} ~\sech^2(ky) \ , 
\end{equation}
which can be solved to give
\begin{equation}\label{solN2}
\phi(y)= \pm\frac12\sqrt{\frac{6B(6+B)(2+5B)}{8+16B+5B^2}}~\tanh(k y)\,.
\end{equation}
Back to equation \eqref{eq1}, we note that the standard solution is obtained with $\alpha=0$; it is given by
\begin{equation}\label{trivsol}
\phi(y)=\sqrt{\frac{3B}{2}} ~\arcsin\Big[\tanh(ky)\Big]\,.
\end{equation}
Additionally, with the choice \eqref{warpfact} can write the energy density \eqref{enerdensiM1} as
\ben\label{enerdensN2}
\rho(y)\!\!\!&=&\!\!\!-B^2 k^2 (3 + 20 \alpha B^2 k^2)~\sech^{2B}(ky)+\nn
\!\!\!&&\!\!\!+a_1\, (1\!+\!2 B)~\sech^{2B+2}(ky)-\nn
\!\!\!&&\!\!\!-a_2\, (1\!+\!B)  ~\sech^{2B+4}(ky)\,,
\een
where $a_{1}$ and $a_{2}$ are given by \eqref{a12}. An important feature of this model is the presence of the split of energy density. This aspect was first pointed out in \cite{bRR}, and it appears for
\begin{equation}
\alpha\to\alpha_s\equiv \frac{9 B+3}{8k^2 \left(49 B^2+60 B+16\right) }\,.
\end{equation}
This and \eqref{range} impose that $-5.68<B<-2.56$ or $B>-0.25$. In Fig.~\ref{fig1} we depict the behavior of the energy density for several values of $\alpha$, using $k=1$, $B=1$ and $\alpha_s=3/250$.

Though for generic values of the parameters it is not always possible to get the scalar potential as a function of the field $\phi$ explicitly, for the particular solution \eqref{solN2} it is indeed possible. In this case, 
\ben
V(\phi)&=&\frac{3k^2 B}{4}\left(\frac{5B^2+24B+12 }{5 B^2+16
   B+8}\right)-k^2(1+2B)\phi^2+\nn
&&+\frac{k^2}{3}\left(\frac{(1+2B)(8 + 16 B + 5 B^2)}{B(6+B)(2+5B)}\right)\phi^4.
\een
This potential engenders spontaneous symmetry breaking and supports the kink-like structure already shown in \eqref{solN2}.

\subsection{Second Example}

Let us now analyze the model with cubic term in the Ricci scalar, that is,
$F(R)=R+\alpha R^3$. In this case, the above equations \eqref{eqphiPot} become
\bes\label{E_equations_M2}
\ben
\phi^{\prime2}\!\!\!&=&\!\!\!-\frac{3}{2}A^{\prime\prime}\!\!+\!24\alpha  \Big(25 A^{\prime 4}\!A^{\prime\prime}\!\!-\!320A^{\prime 2}\!A^{\prime\prime 2}\!\!-\!80 A^{\prime3}\!A^{\prime\prime\prime}\!\!-\nn
&&\!\!\!-52A^{\prime\prime3}-112A^{\prime}A^{\prime\prime}A^{\prime\prime\prime}-8A^{\prime\prime\prime2}-\nn
&&\!\!\!-20A^{\prime2}A^{\prime\prime\prime\prime}-8A^{\prime\prime}A^{\prime\prime\prime\prime}\Big)\,,\label{phiprime_M2} \\
V(\phi) \!\!\!&=&\!\!\!-\frac34 A^{\prime\prime}\!\!-\!3A^{\prime2}\!\!+\!4\alpha\Big(100A^{\prime6}\!\!-\!1845A^{\prime4}\!A^{\prime\prime}\!\!-\!92A^{\prime\prime 3}\!-\nn
\!\!\!&&\!\!\!-1584A^{\prime2}A^{\prime\prime2}-720A^{\prime3}A^{\prime\prime\prime}-528A^{\prime}A^{\prime\prime}A^{\prime\prime\prime}-\nn
\!\!\!&&\!\!\!-24A^{\prime\prime\prime2}-60A^{\prime2}A^{\prime\prime\prime\prime}-24A^{\prime\prime}A^{\prime\prime\prime\prime}\Big)\,,
\label{potential_M2}\een
\ees 
and the energy density can be written as
\ben\label{ener_densi_M2}
\rho(y)\!\!\!&=&\!\!\! -e^{2A}\Bigl(3A^{\prime2}+\frac32A^{\prime\prime}-8\alpha \Big(50A^{\prime6}\!-\!885A^{\prime4}A^{\prime\prime}\!-\nn
\!\!\!&&\!\!\!-1272A^{\prime2}\!A^{\prime\prime2}\!\!-\!124A^{\prime \prime3}\!\!-\!480A^{\prime3}\!A^{\prime\prime\prime}\!\!-\!432A^{\prime}\!A^{\prime\prime}\!A^{\prime\prime\prime}\!\!-\!\nn
\!\!\!&&\!\!\!-24A^{\prime\prime\prime 2}-60A^{\prime2}A^{\prime\prime\prime\prime}-24A^{\prime\prime}A^{\prime\prime\prime\prime}\Big)\Bigr)\,.
\een

\begin{figure}[t]
\begin{center}
\includegraphics[scale=0.7]{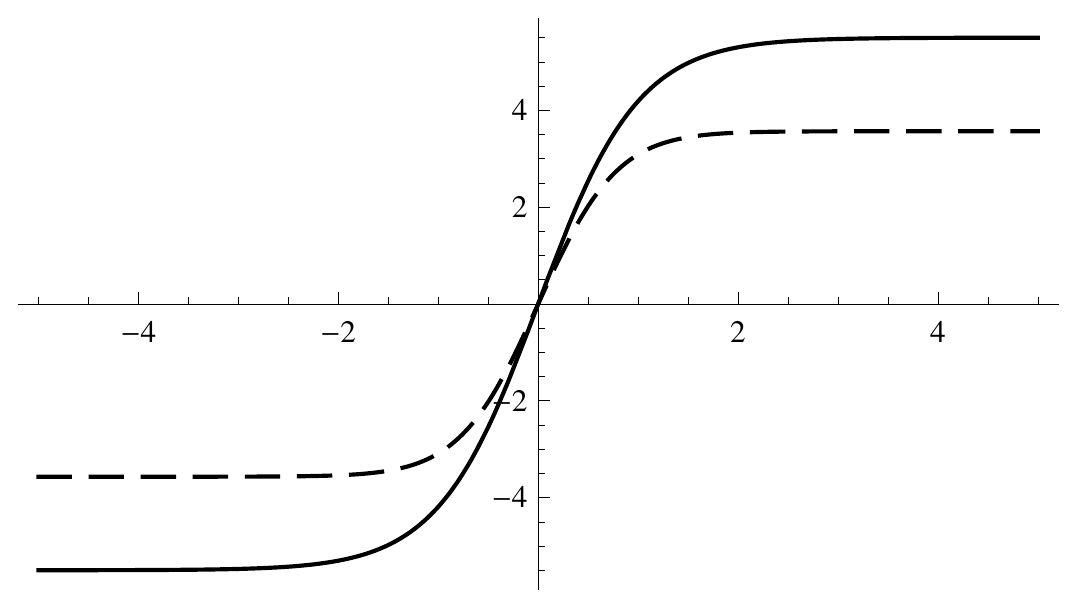}
\end{center}
\vspace{-0.5cm}
\caption{\small{The solution \eqref{sol1n3} (solid line) and \eqref{sol2n3} (dashed line), depicted for $k=1$. \label{fig2}}}
\end{figure}

\begin{figure}[h]
\begin{center}
\includegraphics[scale=0.72]{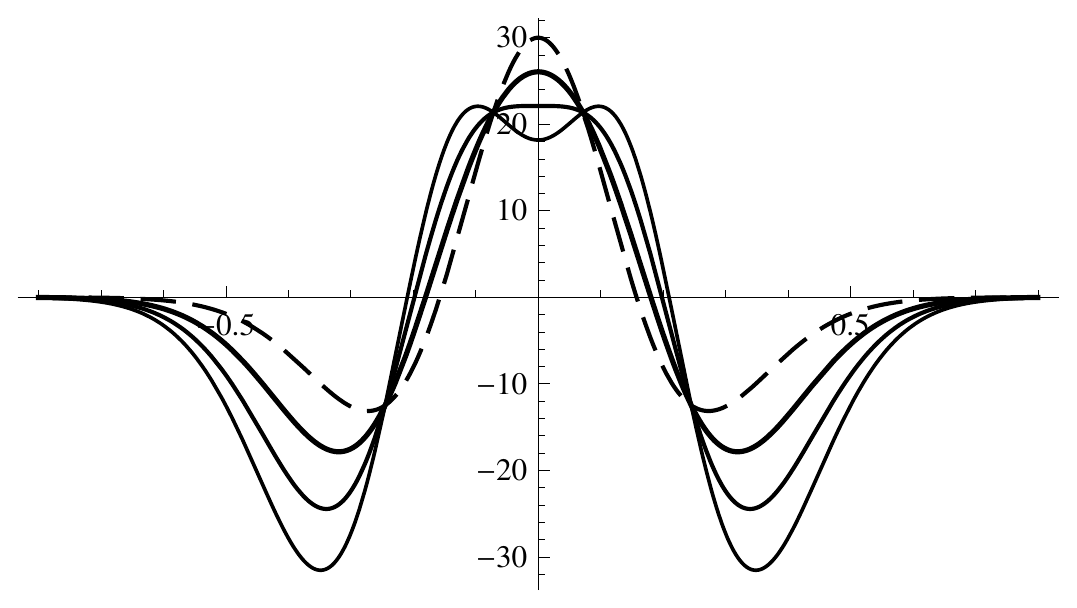}
\end{center}
\vspace{-0.5cm}
\caption{\small{The energy density \eqref{enerdensN3} depicted for $k=1$, $B=20$ and $\alpha=0$ (dashed line), $\alpha=\alpha_{3s}/2$ (thicker line), $\alpha=\alpha_{3s}$ (thick line) and $\alpha=3\alpha_{3s}/2$ (thinner line).\label{fig3}}}
\end{figure}

As we can see, the equation for the scalar field depends on the fourth derivative of the warp function, and suggests that we can use the above procedure to obtain exact solutions. We then suppose that the warp function is again given by \eqref{warpfact}. With this, the differential equation for the scalar field can be written as
\begin{equation}\label{difN3}
\phi^{\prime 2}=b_1\sech^2(k y)-b_2\sech^4(k y)+b_3\sech^6(k y)\,,
\end{equation}
where the constants $b_1$, $b_2$ and $b_3$ are given by
\bes
\ben
b_1&=&\frac{3 B k^2}{2}+120 \alpha B^3 k^6 \left(16 +32B-5 B^2\right)\,, \\
b_2&=&48\alpha B^2k^6 (2 + 5 B) (16+66B-5B^2)\,, \\
b_3&=&24\alpha B^2k^6 (20-B)  (2 + 5 B)^2 \,.
\een
\ees
In this new case, there exist several possibilities to find analytical solutions, including \eqref{trivsol} when $\alpha=0$, as expected. Taking $B=20$ we see that $b_3=0$, which provides the first non-trivial simplification. Assuming now that $\alpha$ is such that $b_1=0$ (ie, $\alpha=1/(43008000 k^4)$) we can write the differential equation \eqref{difN3} as
\begin{equation}
\phi^\prime=\pm k \sqrt{\frac{4233}{140}} ~\sech^2(ky)\,.
\end{equation}
This differential equation admits a solution of the form
\begin{equation}\label{sol1n3}
\phi(y)=\pm \sqrt{\frac{4233}{140}}\tanh(k y)\,,
\end{equation}
which has the same behavior as the solution \eqref{solN2} obtained in the previous model. In this case, the potential as a function of $\phi$ is given by
\be
V(\phi)=\frac{12659}{840}k^2-41k^2\phi^2+\frac{2870}{4233}k^2\phi^4.
\ee
Another possibility is obtained if $B=(33 + \sqrt{1169})/5$ and $\alpha$ is such that $b_1=0$. In this case  the differential equation \eqref{difN3} becomes
\begin{equation}
\phi^\prime=\pm \sqrt{b_3} ~\sech^3(ky) \ , 
\end{equation}
which leads to the new solution
\ben\label{sol2n3}
\phi(y)\!\!\!&=&\!\!\!\pm \sqrt{\frac{21}{340} \!\left(163\!+\!5\sqrt{1169}\right)}~\Big[\!\arctan\!\Big(\!\tanh(ky/2)\Big) \!\!+\nn
\!\!\!&&\!\!\!+\frac12 \sech(k y) \tanh(k y)\Big]\,.
\een
In this case, we cannot get the scalar potential in terms of the field $\phi$ explicitly. In Fig.~\ref{fig2} we illustrate the behavior of the solutions \eqref{sol1n3} (solid line) and \eqref{sol2n3} (dashed line), depicted for $k=1$.
In the general case, we can express the energy density as
\ben\label{enerdensN3}
\rho(y)\!\!\!&=&\!\!\!-B^2 k^2 (3 - 400 \alpha B^4 k^4)~\sech^{2B}(ky)+\nn
\!\!\!&&\!\!\!+b_1 (1\!+\!2 B)~\sech^{2B+2}(ky)+\nn
\!\!\!&&\!\!\!+b_2 (1\!+\!B)  ~\sech^{2B+4}(ky)+\nn
\!\!\!&&\!\!\!+\frac{b_3}{3} (3\!+\!2B)  ~\sech^{2B+6}(ky)\,.
\een
Again, we can show that the energy density begins to split when
\begin{equation}
\alpha\to\alpha_{3s}\equiv-\frac{9 B+3}{64 Bk^4 \left(349 B^2+351 B+84 \right) }\,.
\end{equation}
We see that in this new situation the splitting appears for $\alpha$ negative. In Fig.~\ref{fig3} we show the behavior of the energy density near the splitting. Here we are using $k=1$ and $B=20$,  with $\alpha_{3s}=-183/187781120$.

\subsection{Third Example}

\begin{figure}[t]
\begin{center}
\includegraphics[scale=0.64]{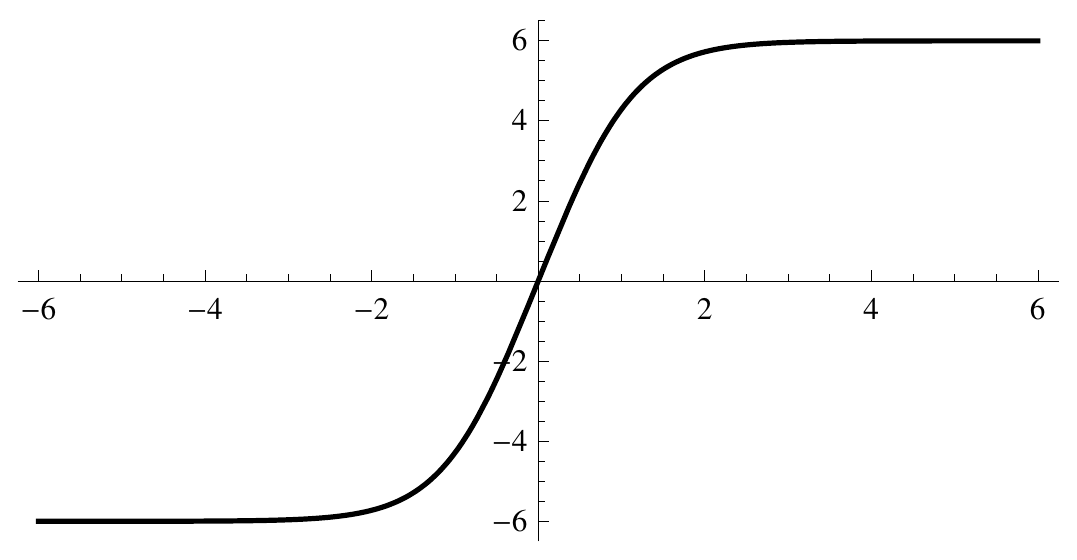}
\end{center}
\vspace{-0.5cm}
\caption{\small{The solutions \eqref{N4sol3}, depicted for $k=1$.\label{fig4a}}}
\end{figure}

Let us now study the case with $F(R)=R+\alpha R^4$. Here the differential equation \eqref{eqphi} for the scalar field becomes
\ben
\phi^{\prime2}\!\!\!&=&\!\!\!-\frac32 A^{\prime\prime}-6\alpha A^{\prime\prime}\left(20A^{\prime2}+8A^{\prime \prime}\right)^{3} +\nn
\!\!\!&&\!\!\!+48\alpha \!\left(5A^{\prime2}\!A^{\prime\prime}\!\!-\!5A^{\prime\prime2}\!-\!4A^{\prime}\!A^{\prime\prime\prime}\!\!-\!A^{\prime\prime\prime\prime}\right) \times \nn
\!\!\!&&\!\!\!\times \left(20A^{\prime2}+8A^{\prime \prime}\right)^{2} \!-768\alpha \times\nn
\!\!\!&&\!\!\! \times\left (5A^\prime A^{\prime\prime} + A^{\prime\prime\prime}\right)^2\left(20A^{\prime2}+8A^{\prime \prime}\right)\,.
\een
Thus, using equation \eqref{warpfact}, we find that
\ben
\phi^{\prime2}&=&c_1~\sech^{2}(ky)+c_2~\sech^{4}(ky)+\nn
&&+c_3~\sech^{6}(ky)+c_4~\sech^{8}(ky)\,,\label{difN4}
\een
where the constants are given by
\bes
\ben
c_1\!\!\!&=&\!\!\! \frac{3 B k^2}{2}\!+\!9600 \alpha k^8 B^5 \left(8\!+\!16B\!-\!5B^2\right) \,,\\
c_2\!\!\!&=&\!\!\!1920 \alpha k^8 B^4 (2\!+\!5B) \left(15 B^2\!-\!102B\!-\!32 \right)  \,,\\
c_3\!\!\!&=&\!\!\! 1152 \alpha k^8 B^3 (2\!+\!5B)^2 \left(8 + 52 B - 5 B^2\right)\,,\\
c_4\!\!\!&=&\!\!\! 384 \alpha k^8 B^3 (2\!+\!5B)^3(B-14)\,.
\een
\ees
In this case we suppose that
\be
 \alpha\to\alpha_1=1/(6400 B^4k^6(5B^2-16B-8)\,,
 \ee
then $c_1=0$. Also, we take $B$ as one of the possible solutions of the following equation
\begin{equation}\label{restriction}
-\frac{3\left(8 + 52 B - 5 B^2\right)}{2(B-14)}=\pm\sqrt{\frac{5 B \left(15 B^2\!-\!102B\!-\!32 \right)}{(B-14)}}\,.
\end{equation}
With this restriction, the differential equation \eqref{difN4} becomes
\begin{equation}\label{difN4simpl}
\phi^{\prime}=\pm\sqrt{\frac{3k^2(2\!+\!5B)^3(B\!-\!14)}{50B(5B^2\!-\!16B-8)}} \sech^{2}(ky)\Big(\varepsilon+\sech^{2}(ky)\Big)\,,
\end{equation}
where
\be \varepsilon=\frac{3\left(8+52B-5B^2\right)}{2(2+5B)(B-14)}\,.
\ee
For equation \eqref{difN4simpl} to be true, we must impose that $c_4>0$, with $\alpha=\alpha_1$, which causes $\!8/5\!-\!2\sqrt{26}/5<B\!<\!-2/5\!$, or $0\!<B<\!8/5\!+\!2\sqrt{26}/5$, or $B>14$.
Solving the differential equation \eqref{difN4simpl} we obtain, for the allowed values of $B$, the solution 
\be\label{N4sol3}
\phi(y)\!\!=\!\!\!\pm\left(5.99 - 0.87\, \sech^2(k y)\right)\! \tanh(ky)\,,
\ee
for $B=17.448$. It is depicted in Fig.~\ref{fig4a}.

The energy density can be written as
\ben\label{ener_densi_M3}
\rho(y)\!\!\!&=&\!\!\! -3B^2 k^2 (1-8000 \alpha B^6 k^6)~\sech^{2B}(ky)+\nn
\!\!\!&&\!\!\!+c_1 (1\!+\!2 B)~\sech^{2B+2}(ky)+\nn
\!\!\!&&\!\!\!+c_2 (1\!+\!B)  ~\sech^{2B+4}(ky)+\nn
\!\!\!&&\!\!\!+\frac{c_3}{3} (3\!+\!2B)  ~\sech^{2B+6}(ky)+\nn
\!\!\!&&\!\!\!+\frac{c_4}{2} (2\!+\!B)  ~\sech^{2B+8}(ky)\,.
\een
We can also see that in this case the energy density starts to split for $\alpha$ given by
\begin{equation}
\alpha\to\alpha_{4s}\equiv \frac{1+3 B}{1024 B^2k^6(40+176B+189B^2)}\,.
\end{equation}
 
\begin{figure}[!htb]
\begin{center}
\includegraphics[scale=0.7]{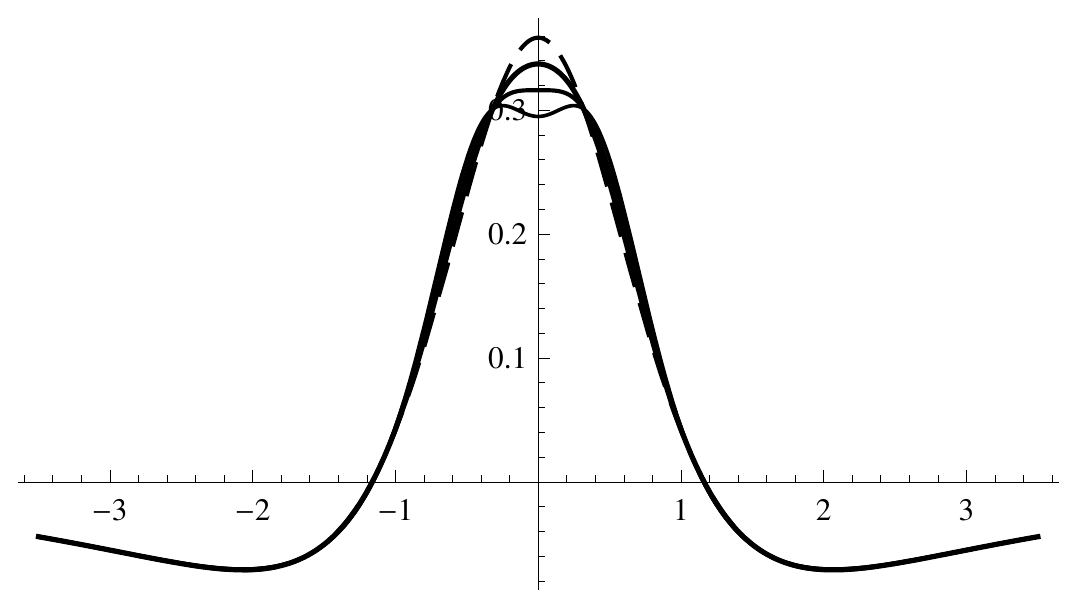}
\end{center}
\vspace{-0.5cm}
\caption{\small{Energy density \eqref{ener_densi_M3} depicted for $k=1$, $B=0.239$ and $\alpha=0$ (dashed line), $\alpha=\alpha_{4s}/2$ (thicker line), $\alpha=\alpha_{4s}$ (thick line) and $\alpha=3\alpha_{4s}/2$ (thinner line).\label{fig5}}}
\end{figure}

To see how the splitting appears, in Fig.~\ref{fig5} we depict the energy density \eqref{ener_densi_M3} for some values of $\alpha$,
with $k=1$, $B=0.239$ and $\alpha_{4s}=0.00032$.

The above investigations show that the procedure is very efficient to obtain analytical solutions for $n=2,3,$ and $4$, and this is important progress. However, we note that when $n$ increases, the number of free parameters becomes severely constrained. In particular, if $n$ is greater then $4$ this makes it harder to find analytical solutions. We can circumvent the problem relying on numerical investigation, an issue that is out of the scope of the current work. 
​​

To complete our study, we must determine if the modifications of gravity  introduced in the above examples can contribute to destabilize the brane. This is done in the next section, where we investigate their linear stability in the gravity sector (tensorial modes).

\section{Stability}

The investigation of linear stability of the generalized braneworld models presented above can be carried out in different levels of complexity, depending on whether one is interested in the scalar field sector and/or in the gravitational side. Here, for simplicity, we focus on purely tensorial perturbations which, as is well known, are gauge independent. We can thus assume that the brane remains static (Gaussian normal coordinates) and write the perturbed line element as (here $x$ stands for a vector in the $(3,1)$ space-time)
\ben
ds^2=e^{2A(y)}\Big[\eta_{\mu\nu}+h_{\mu\nu}(x, y)\Big] dx^\mu dx^\nu -dy^2\,.\label{eq13}
\een
Up to first-order in the tensorial perturbation $h_{\mu\nu}$,  the field equations yield
\ben
\!\!&&\!\!\Big[\frac12h_{\mu\nu}^{\prime\prime}+2A^\prime h_{\mu\nu}^\prime-\frac12e^{-2A}\Box^{(4)} h_{\mu\nu}+\nn
\!\!&&\!\!+\frac12e^{-2A}\Big(\partial_\mu\partial^\alpha h_{\alpha\nu}\!+\!\partial_\nu\partial^\alpha h_{\alpha\mu}\!-\!\partial_\mu\partial_\nu h\Big)\Big]F_R\!+\!\frac12h_{\mu\nu}^\prime F_R^\prime\!-
\nn
\!\!&&\!\!-e^{-2A}\partial_\mu\partial_\nu P(x,y)= \Big\{3A^\prime P^\prime(x,y)+P^{\prime\prime}(x,y)-\nn
\!\!&&\!\!-e^{-2A}\Box^{(4)} P(x,y)+\Big(A^\prime h+\frac12 h^\prime\Big)F_R^\prime+\frac12A^\prime h^\prime+\nn
\!\!&&\!\!+2V_\phi\xi\!+\!2\phi^{\prime} \xi^{\prime}\!-\!\Big[4A^{\prime 2}\!+\!A^{\prime\prime}\!-\!\frac12\frac{F_R}{F_{RR}}\Big]P(x,y)\Big\}\eta_{\mu\nu} \,,~~~~~~\label{eq15}
\een
where $h=h^{\mu}{}_\mu$ and the function $P(x,y)$ is defined as
\ben
P(x,y)\!=\!\Big[h^{\prime\prime}\!\!+\!5A^\prime h^\prime\!+\!e^{-2A}\!\left(\partial^\mu\partial^\nu h_{\mu\nu}\!-\!\Box^{(4)}h\right)\!\Big]\!F_{RR}.\;\;\label{eq16}
\een
As usual, we consider the perturbations to be transverse and traceless, i.e., we take
\begin{equation}\label{eq18}
\partial^\mu h_{\mu\nu}=0 \mbox{\;\;\;\;\; and \;\;\;\;\;}h=0\ ,
\end{equation}
which reduces Eq.~\eqref{eq15} to the form
\ben
&&\Big(\!-\!\partial_y^2\! -\!4A^\prime \partial_y\!+\!e^{-2A}\Box^{(4)} \!-\!\frac{F_R^\prime}{F_R}\partial_y\!\Big)h_{\mu\nu}=0 \,.\label{eq19}
\een
We now introduce a new coordinate $z$ via the transformation 
\begin{equation}\label{eq20a}
dz=e^{-A(y)}dy\ , 
\end{equation}
such that the line element is now conformally flat, and decompose the modes in the form
\ben\label{eq20}
h_{\mu\nu}(x,z)=e^{-ip\cdot x}e^{-3A(z)/2 }F_R^{-1/2}\bar h_{\mu\nu}(z) \,.
\een
By doing this,  the 4-dimensional components of $h_{\mu\nu}$ obey the Klein-Gordon equation and 
the metric fluctuations of the brane satisfy  the Schr\"{o}dinger-like equation
\ben\label{eq21}
\left[-\frac{d^2}{dz^2} + U(z) \right]\bar h_{\mu\nu} = p^2 \bar h_{\mu\nu}\,,
\een
where 
\ben\label{eq22}
U(z)&=&\frac{9}{4} A_z^2 + \frac32 A_{zz}+\frac32A_z \frac{d(\ln F_R)}{dz}-\nn
&&-\frac14 \Big(\frac{d(\ln F_R)}{dz}\Big)^2+\frac1{2F_R}\frac{d^2 F_R}{dz^2}\,.
\een
The stability of the above modes can be established by noting that one can write
\be
-\frac{d^2}{dz^2} + U(z) =S^\dagger S\,,\label{eq23}
\ee
where
\be
S=-\frac{d}{dz}+\frac32 A_z+\frac12 \frac{d(\ln F_R)}{dz}\,.\label{eq24}
\ee
This decomposition puts forward that the operator $-\partial_z^2 + U(z)$ is non-negative, which guarantees that the gravity sector is linearly stable. We note that the stability behavior in the gravity sector only depends on the warp factor since $R$ is a function of $A(z)$, explicitly given by $R=4e^{-2A(z)}(3 A_z^2+2A_{zz})$. In addition, the zero mode can be written as
\begin{equation}\label{graundstatFR0}
\eta_0(z)=N_0~e^{3A(z)/2} F_R^{1/2}\,.
\end{equation}

For the polynomial family of gravity Lagrangians defined in equation \eqref{eqFR},  the potential of stability \eqref{eq22} can be written in the specific form
\ben\label{stabilPot}
U(z)\!\!\!&=&\!\!\!\frac{9}{4} A_z^2 + \frac32 A_{zz}-\frac{\alpha n(n-1)}{3 A_z^2+2A_{zz}}\Big(\frac{R^{n-1}}{1+\alpha nR^{n-1}}\Big) \times\nn 
\!\!\!&&\!\!\!\times\Big[3A_z^4 +8A_z^2A_{zz}-A_{zz}^2-2A_{z}A_{zzz}-A_{zzzz}\Big]+\nn
&&+\frac{\alpha n(n-1)R^{n-1}}{1+\alpha nR^{n-1}} \Big(\frac{3A_z^3-A_zA_{zz}-A_{zzz}}{3 A_z^2+2A_{zz}}\Big)^2 \times\nn
\!\!\!&&\!\!\!\times\Big[2(n-2)-\frac{\alpha n(n-1)R^{n-1}}{1+\alpha nR^{n-1}} \Big]\ ,
\een
whose zero mode is readily seen from \eqref{graundstatFR0} and reads
\begin{equation}\label{graundstatFR}
\eta_0(z)\!=\!N_0 e^{3A(z)/2}\sqrt{1\!+\!\alpha n R^{n-1} } \,.
\end{equation}
The next step is to make the change of variable \eqref{eq20a} for the choice of warp function introduced in \eqref{warpfact}. This can be done in a general way so that it allows us to obtain the following relationship between the $z$ and $y$ coordinates
\begin{equation}\label{zinvers}
z=\frac{\sinh(ky)}{k}~{}_2F_1\Big(\frac12, \frac{1\!-\!B}{2}, \frac32, -\sinh^2(ky)\Big) \,,
\end{equation}
where ${}_2F_1$ is hypergeometric function. Note that this equation can not be inverted analytically for arbitrary $B$, though it is possible for some particular choices. For instance, if $B=1$ we obtain $y=\arcsinh(kz)/k$. 

With all these elements, we are ready to investigate the linear stability of the polynomial models presented  above.

\subsection{Case $n=2$}

For the first example, involving a quadratic term in the Ricci scalar, the stability potential \eqref{stabilPot} can be written as
\ben
U(z)\!\!\!&=&\!\!\!\frac{9}{4} A_z^2 + \frac32 A_{zz}-\Big(\frac{8 \alpha }{e^{2A(z)}+8\alpha (3 A_z^2+2A_{zz})}\Big) \times\nn 
\!\!\!&&\!\!\!\times\Big[3A_z^4 +8A_z^2A_{zz}-A_{zz}^2-2A_{z}A_{zzz}-A_{zzzz}\Big]-\nn
&&-64\alpha^2\Big(\frac{3A_z^3-A_zA_{zz}-A_{zzz}}{ e^{2A(z)}+8\alpha (3 A_z^2+2A_{zz})}\Big)^2\,.
\een

\begin{figure}[t]
\begin{center}
\includegraphics[scale=0.7]{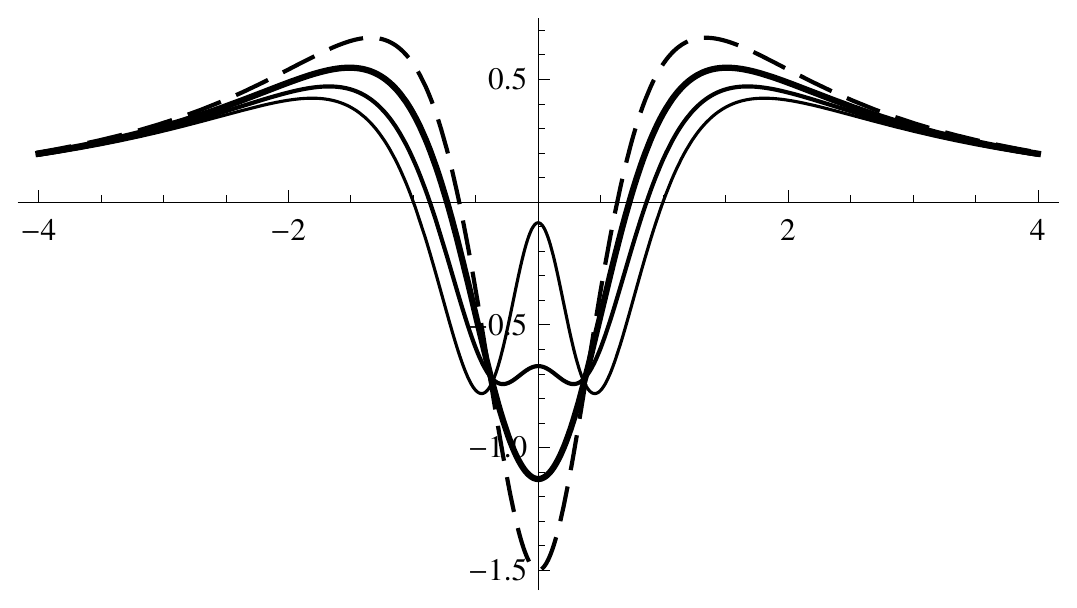}
\end{center}
\vspace{-0.7cm}
\caption{\small{Stability potential \eqref{eq40} depicted for $k=1$ and and $\alpha=0$ (dashed line) , $\alpha=\alpha_s/2$ (thicker line), $\alpha=\alpha_s$ (thick line) and $\alpha=3\alpha_s/2$ (thinner line) with $\alpha_s=3/250$.\label{fig6}}}
\end{figure}

\begin{figure}[t]
\begin{center}
\includegraphics[scale=0.7]{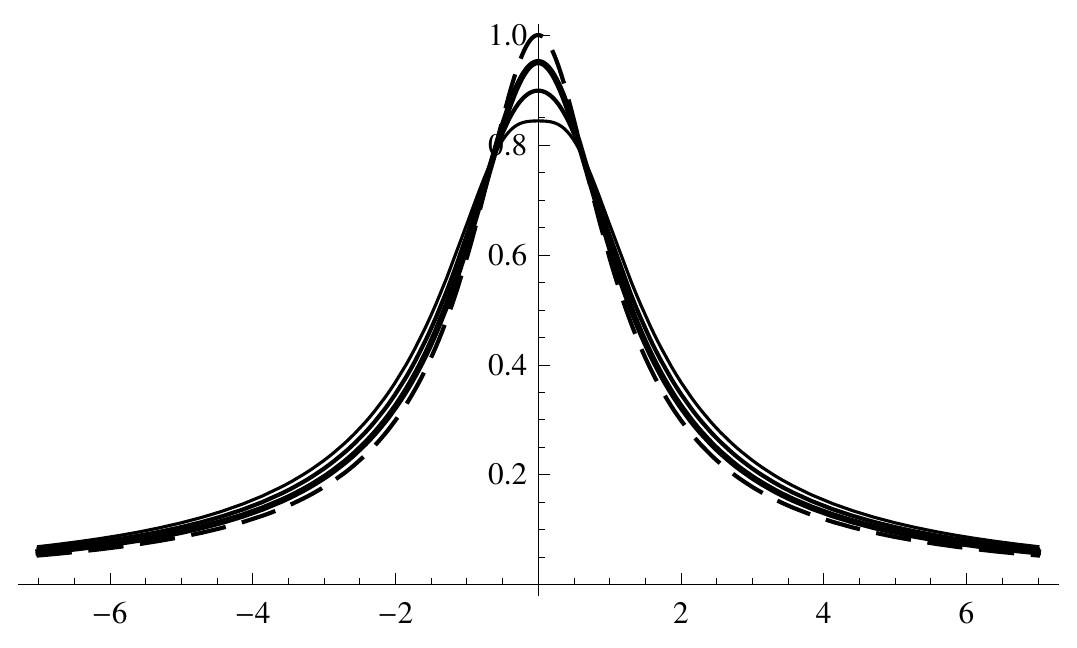}
\end{center}
\vspace{-0.7cm}
\caption{\small{Zero mode \eqref{estFunN2} depicted for $k=1$ and and $\alpha=0$ (dashed line) , $\alpha=\alpha_s/2$ (thicker line), $\alpha=\alpha_s$ (thick line) and $\alpha=3\alpha_s/2$ (thinner line) with $\alpha_s=3/250$.\label{fig7}}}
\end{figure}

In this case, the parameter $B$ can take any positive value and, for simplicity, we set it to $B=1$. This choice has the clear advantage of allowing to perform analytically the inversion of equation \eqref{zinvers}, so that the stability potential becomes
\ben\label{eq40}
U( z)\!\!\!&=&\!\!\!-\frac{3 k^2}{4}\frac{2-5 k^2z^2}{\left(1+k^2 z^2\right)^2}-\nn
\!\!\!&&\!\!\!+\frac{56\alpha k^4 \left(1-6k^2 z^2\right)}{\left(1\!+\!k^2 z^2\right)^2 
\left[1\!+\! k^2z^2\!-\!8 \alpha k^2 \left(2\!-\!5 k^2 z^2\right)\right]}-~~~~\nn
\!\!\!&&\!\!\!-\frac{3136 \alpha^2 k^8 z^2}{\left(1\!+\!k^2 z^2\right)^2 \left[1\!+\! k^2z^2\!-\!8 \alpha k^2 \left(2\!-\!5 k^2 z^2\right)\right]^2}\,.~~
\een
In addition, the zero mode is
\begin{equation}\label{estFunN2}
\eta_0(z)=\frac{\sqrt{1-8\alpha k^2\left(\frac{2-5 k^2z^2}{1+k^2 z^2}\right)}}{\left(1+k^2 z^2\right)^{3/4}} \,.
\end{equation}
The potential of stability \eqref{eq40} and the zero mode \eqref{estFunN2} are depicted in Fig.~\ref{fig6} and Fig.~\ref{fig7}, respectively. In both cases we used values of $\alpha$ around the splitting value. It should be noted that as the splitting of the energy density occurs, the form of the potential at $y=0$ changes dramatically from a deep minimum to a local maximum. As a result, the ground state becomes less localized, with the top of the wave function being lower and its width larger. 


\subsection{Case  $n=3$}

\begin{figure}[!htb]
\begin{center}
\includegraphics[scale=0.7]{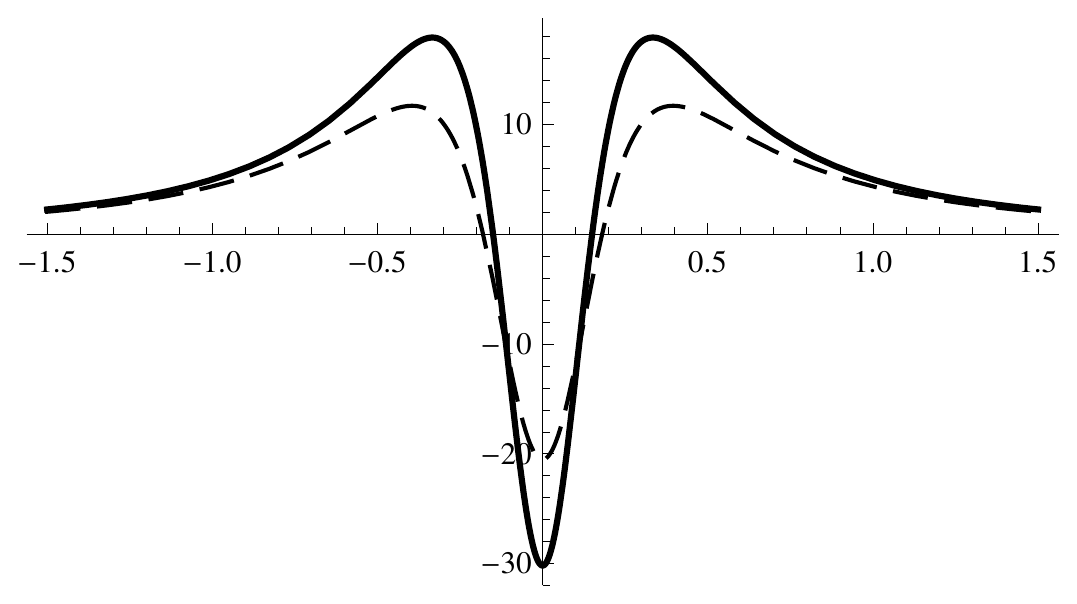}
\end{center}
\vspace{-0.7cm}
\caption{\small{Stability potential \eqref{stabilPotN3} depicted numerically for $k=1$, $\alpha=1/43008000$, and $B=20$  (solid line) and  $B=(33+\sqrt{1169})/5$ (dashed line). \label{fig8}}}
\end{figure}

\begin{figure}[!htb]
\begin{center}
\includegraphics[scale=0.7]{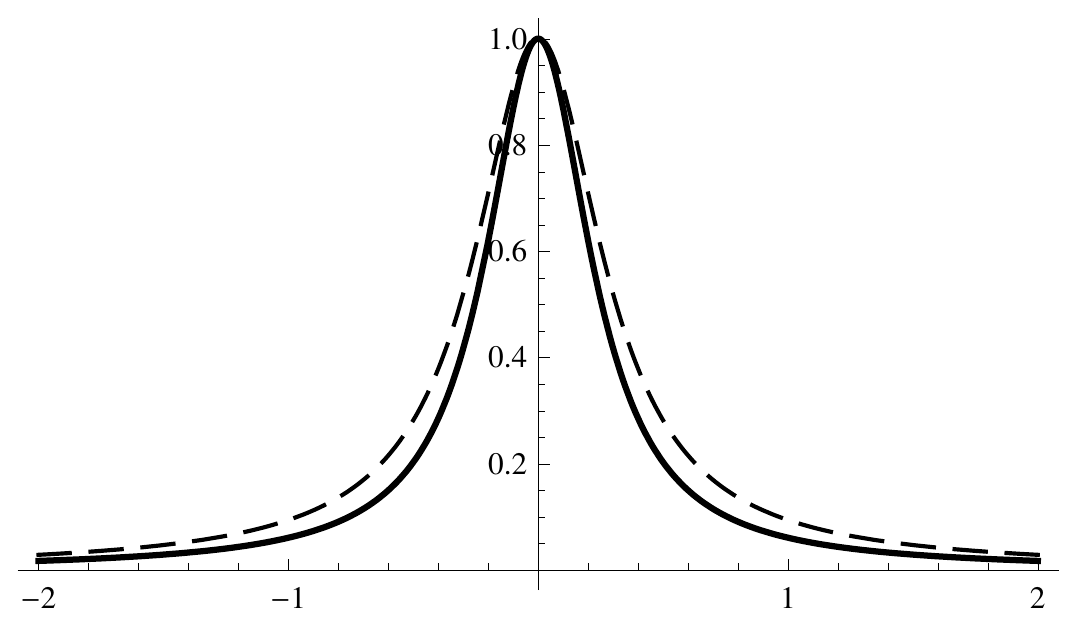}
\end{center}
\vspace{-0.7cm}
\caption{\small{Numerical representation of the zero mode \eqref{estFunN3} for $k=1$, $\alpha=1/43008000$, and $B=20$  (solid line) and  $B=(33+\sqrt{1169})/5$ (dashed line).\label{fig9}}}
\end{figure}

Let us now analyze the model with cubic corrections in the Lagrangian. In this case the stability potential \eqref{stabilPot} becomes
\ben\label{stabilPotN3}
U(z)\!\!\!&=&\!\!\!\frac{9}{4} A_z^2 \!+\! \frac32 A_{zz}\!-\!96\alpha \Big[\!\frac{3 A_z^2\!+\!2A_{zz}}{e^{4A(z)}\!+\!48\alpha (3 A_z^2\!+\!2A_{zz})^2}\Big]\!\!\times\nn
\!\!\!&&\!\!\!\times \Big[3A_z^4 \!+\!8A_z^2A_{zz}\!-\!A_{zz}^2\!-\!2A_{z}A_{zzz}\!-\!A_{zzzz}\Big]\!+\!\nn
\!\!\!&&\!\!\!+192\alpha e^{4A(z)}  \Big(\frac{3A_z^3-A_zA_{zz}-A_{zzz}}{e^{4A(z)}\!+\!48\alpha (3 A_z^2\!+\!2A_{zz})^2}\Big)^2\,.
\een
Though we found two exact solutions for this model, the corresponding values of $B$ do not allow for an analytical inversion of equation \eqref{zinvers}. Nonetheless,  the inversion can be carried out numerically, which is done in Figure~\ref{fig8}, using $k=1$, $B=20$  (solid line) and  $B=(33+\sqrt{1169})/5$ (dashed line), with $\alpha=1/43008000$.
For this model, the zero mode \eqref{graundstatFR} can be written as
\begin{equation}\label{estFunN3}
\eta_0(z)= e^{3A(z)/2}\sqrt{1+48\alpha e^{-4A(z)}\Big(3 A_z^2+2A_{zz}\Big)^2}  \ ,
\end{equation}
whose numerical representation appears in Fig~\ref{fig9}. It is apparent from this plot that the localization of the ground state around the brane is tighter for smaller values of $B$, with the potential minimum being deeper and the local maxima higher.

\subsection{Case $n=4$}

\begin{figure}[!htb]
\begin{center}
\includegraphics[scale=0.7]{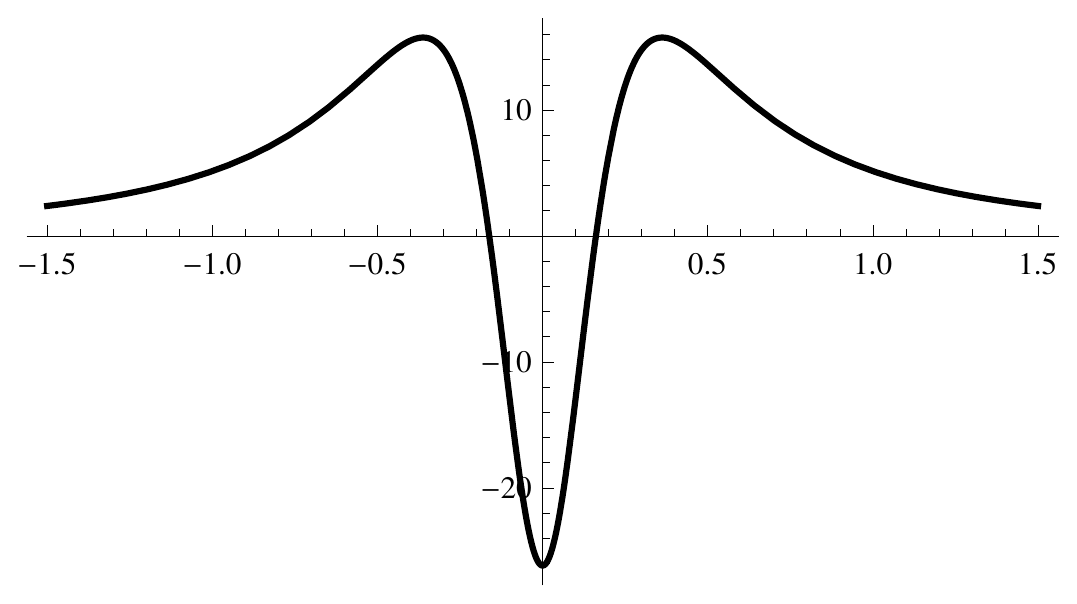}
\end{center}
\vspace{-0.7cm}
\caption{\small{Stability potential \eqref{stabilPotN4}, depicted for $k=1$ and $B\!=\!17.448$ and $\alpha\!=\!1.36\times 10^{-12}$.\label{fig10}}}
\end{figure}

\begin{figure}[!htb]
\begin{center}
\includegraphics[scale=0.7]{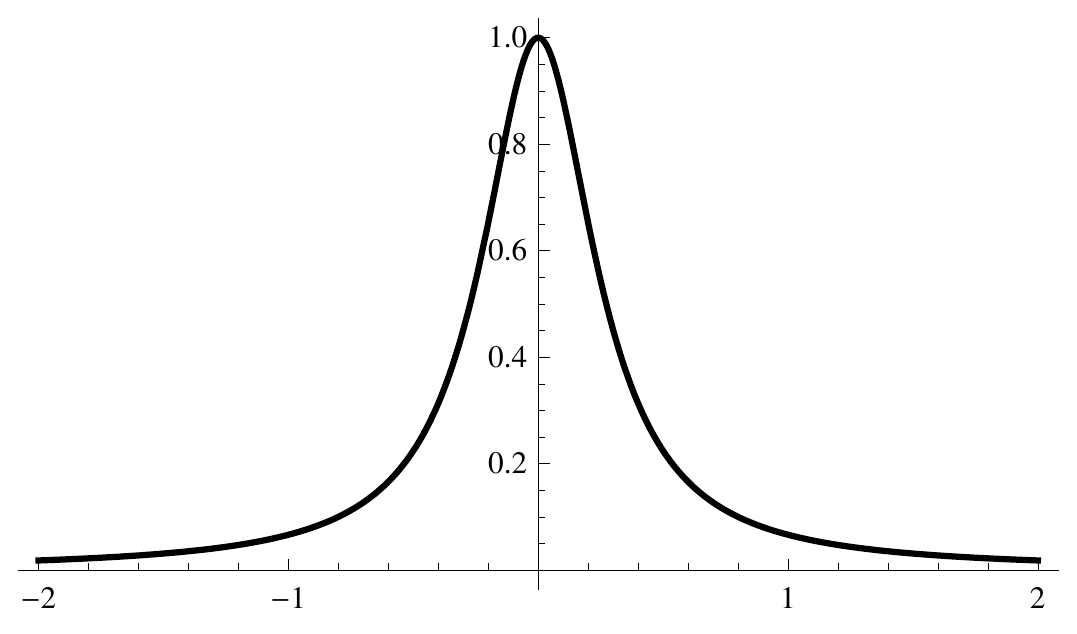}
\end{center}
\vspace{-0.7cm}
\caption{\small{The zero mode \eqref{estFunN4}, depicted for $k=1$, $B=17.448$ and $\alpha\!=\!1.36\times 10^{-12}$.\label{fig11}}}
\end{figure}

For $n=4$, the stability potential \eqref{stabilPot} reads
\ben\label{stabilPotN4}
U(z)\!\!\!&=&\!\!\!\frac{9}{4} A_z^2 + \frac32 A_{zz}-\frac{768\alpha (3 A_z^2+2A_{zz})^2}{e^{6A(z)}+256\alpha (3 A_z^2+2A_{zz})^3}\times\nn 
\!\!\!&&\!\!\!\times\Big[3A_z^4 +8A_z^2A_{zz}-A_{zz}^2-2A_{z}A_{zzz}-A_{zzzz}\Big]+\nn
\!\!\!&&\!\!\!+768\alpha (3 A_z^2\!+\!2A_{zz})\!\Big[\!\frac{(3A_z^3-A_zA_{zz}-A_{zzz})^2}{e^{6A(z)}\!+\!256\alpha (3 A_z^2\!+\!2A_{zz})^3}\!\Big]\!\! \times\nn
\!\!\!&&\!\!\!\times\Big[4-\frac{768\alpha (3 A_z^2+2A_{zz})^3}{e^{6A(z)}+256\alpha (3 A_z^2+2A_{zz})^3} \Big]\,.
\een
Similarly as in the example of $n=3$, the analytical inversion of equation \eqref{zinvers} is not possible. Thus, once again we resort to a computer analysis to study the behavior of the stability potential, which appears in Fig~\ref{fig10}.
For this model, the zero mode takes the form
\begin{equation}\label{estFunN4}
\eta_0(z)\!=\! e^{3A(z)/2}\sqrt{1\!+\!256\alpha e^{-6A(z)}\Big(3 A_z^2\!+\!2A_{zz}\Big)^3}  \ ,
\end{equation}
and is plotted in Fig \ref{fig11}.

As informed in the beginning of the section, we have studied stability against tensorial perturbations. With this, we can then investigate several other related issues, among them the presence of resonances, the Newtonian potential, the entrapment of fermions and gauge fields, and other forms of fluctuations, in particular the addition of scalar perturbations. These problems have been studied in several different contexts, for instance, in \cite{f,c,bR,Dz,Zhong,bd,aq,soda,bRR,liu,gomes,chen,hott,lan,carlos} and in references therein. The point here is that such investigations are in general more involved and would usually rely on numerical study, leading us to issues that are out of the scope of the current work. Instead of following such possibilities, we now broaden the scope of the work extending the scalar field sector, allowing that the source field engenders generalized dynamics, an issue that we deal analytically in the next section.

\section{Extending the scalar sector}

The methodology presented above to obtain exact solutions for brane configurations governed by different gravitational Lagrangians can also be extended to scenarios in which, in addition, the source scalar field has non-standard dynamics. 
For the sake of generality, we can consider models of the form
\be\label{RX}
{\cal S}=\int d^5x \sqrt{|g|}\left(-\frac14 F(R) + G(X) - V(\phi)\right),
\ee
where
\be 
X=\frac12 \nabla_a \phi \nabla^a \phi \ ,
\ee
and $G(X)$ is an arbitrary function of $X$. Modifications of this type in the scalar sector have been considered in the literature in diverse contexts, such as in inflationary models (as a way to produce non-Gaussianities) and in relation with dark energy models. Modifying the form of the scalar sector necessarily has an impact on its solutions as compared to the case of canonical scalar fields. However, as tensorial perturbations do not get mixed with scalar modes, the stability analysis performed in the previous section remains valid \cite{Bazeia,Zhong}. 

In this section, therefore, we focus on the search for analytic solutions in models with non-canonical (or generalized) kinetic term, putting forward in this way the versatility of our approach in different scenarios with modified dynamics. Rather than providing an exhaustive list of models and their solutions, we just point out that successful results can be obtained for models of the form
\be 
F(R)=\sum_{n=1}^N a_n R^n; \;\;\;\;\;G(X)=\sum_{m=1}^M b_m X^m\ ,
\ee
though we will only present results for the illustrative case 
\be 
F(R)=R-\alpha R^2; \;\;\;\;\;G(X)=X-a X^2\ ,
\ee
where $\alpha$ and $a$ are real constants, and $a>0$ is necessary to ensure hyperbolicity \cite{Babichev} of the equation of motion of the scalar field. For other values of $N$ and $M$ the results are very similar. 

\begin{figure}[!htb]
\begin{center}
\includegraphics[scale=0.62]{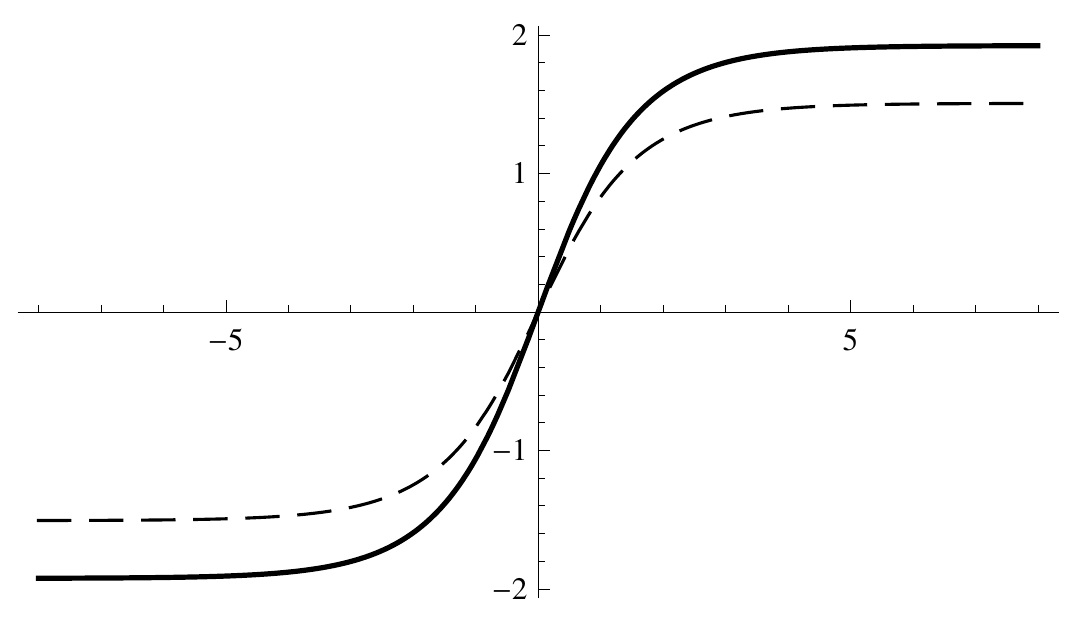}
\end{center}
\caption{\small{The scalar field solution \eqref{eq68}, depicted for $\alpha\!=\!0$ (solid line) and $\alpha\!=\!0.005$ (dashed line).\label{fig12}}}
\end{figure}

\begin{figure}[!htb]
\begin{center}
\includegraphics[scale=0.6]{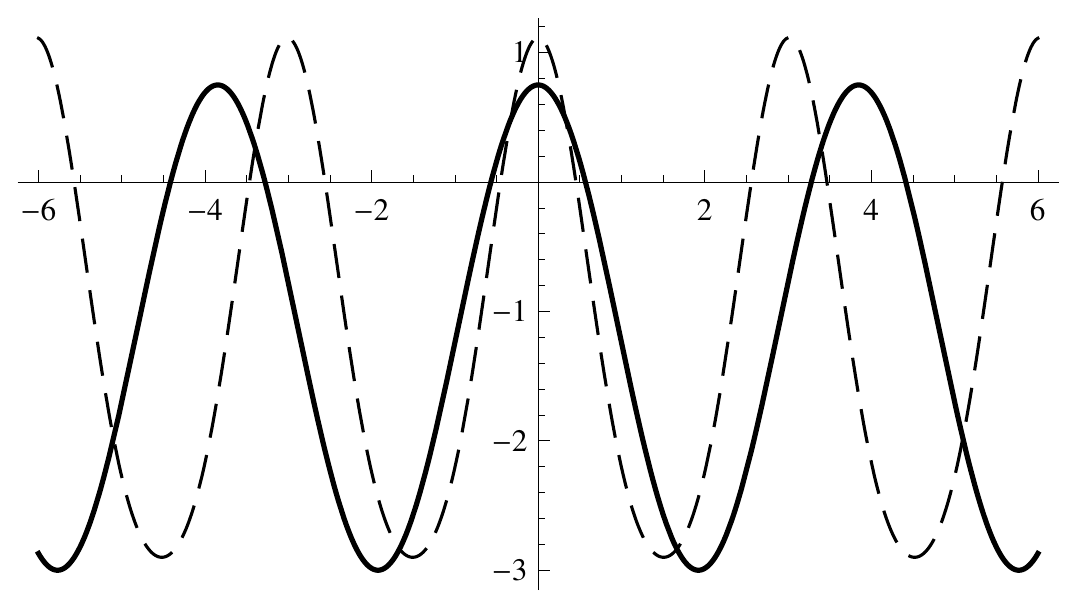}
\end{center}
\caption{\small{The scalar potential \eqref{eq69}, depicted for $\alpha\!=\!0$ (solid line) and $\alpha\!=\!0.005$ (dashed line).\label{fig13}}}
\end{figure}

For the above model with quadratic corrections, the field equations become
\bes\label{eq67}
\ben
\phi^{\prime 2}\!+\!a\phi^{\prime 4} \!\!\!&=&\!\!\!-\frac32A^{\prime \prime}\!\!+\label{eq67.a}\\
\!\!\!&&\!\!\!+4\alpha (2 A^{\prime \prime \prime \prime }\!\!+\!16 A^{\prime\prime 2}\!\!+\!8A^\prime A^{\prime \prime \prime }\!\!+\!5 A^{\prime 2} A^{\prime \prime }),\nn
V(\phi)\!-\!\frac{a}{4}\phi^{\prime 4}\!\!\!&=&\!\!\!-\frac34 A^{\prime\prime}\!\!-\!3A^{\prime2}\!\!+\!\label{eq67.b}\\
\!\!\!&+&\!\!\!2\alpha \!\left(8A^{\prime\prime2}\!\!+\!69A^{\prime2}\!A^{\prime\prime}\!\!+\!\!10A^{\prime4}\!\!+\!\! 24A^{\prime} \!A^{\prime\prime\prime}\!\!+\!2A^{\prime\prime\prime\prime}\right)\!.\nonumber
\een
\ees
Note that if $a=0$ and $\alpha\to-\alpha$ we get back the the equations \eqref{EEM1}, as expected. On the other hand, if $a\neq 0$, the equations involve up to the fourth power of $\phi'$, which makes it more difficult to find solutions. However,  as suggested previously, we can use the standard warp function ansatz \eqref{warpfact} to see how it performs here. With that choice, equation \eqref{eq67.a} turns into
\ben
\phi^{\prime 2}\!+\!a \phi^{\prime 4}\!\!\!&=&\!\!\!\frac12(3\!-\!232\alpha) ~\sech^2(y)\!+\!196\alpha~\sech^4(y)\,,~~~~~\label{eq68a}
\een
where we used $B=1$. This equation can be solved if we assume a suitable choice for the parameters $a$ and $\alpha$. For example, if $a=784\alpha/(3\!-\!232\alpha)^2$, we obtain the following solution
\begin{equation}\label{eq68}
\phi(y)=\sqrt{\frac{3}{2}\!-\!116\alpha}~\arcsin\Big[\tanh(y)\Big]\,.
\end{equation}
For the solution to be real we also must impose that $\alpha\!<\!3/232$. Note that the asymptotic value of the solution is changed to $\phi\!\to\!\pm\bar\phi$ with $\bar\phi\!\equiv\!(\sqrt{2}\pi/4)\sqrt{3\!-\!232\alpha}$ . With the solution \eqref{eq68} we can use equation \eqref{eq67.b} to obtain the scalar potential as
\ben\label{eq69}
V(\phi)\!&=&\!-3+20\alpha+\frac54\left(3-232\alpha\right)\cos^2\Big(\frac{\pi}{2}\frac{\phi}{\bar\phi}\Big)+ \nn
\!&&\!+ ~343\alpha \cos^4\Big(\frac{\pi}{2}\frac{\phi}{\bar\phi}\Big)\,,
\een
where we have $V(\bar\phi)\!=\!-3\!+\!20\alpha$. In Fig.~\ref{fig12} we depict the solution \eqref{eq68} and in Fig.~\ref{fig13} the potential \eqref{eq69}. In the two figures, the value $\alpha\!=\!0$ is represented by the continuous line and $\alpha\!=\!0.005$ by the dashed line. We chose the value $\alpha\!=\!0.005$ because it is inside the allowed interval.

\begin{figure}[t]
\begin{center}
\includegraphics[scale=0.72]{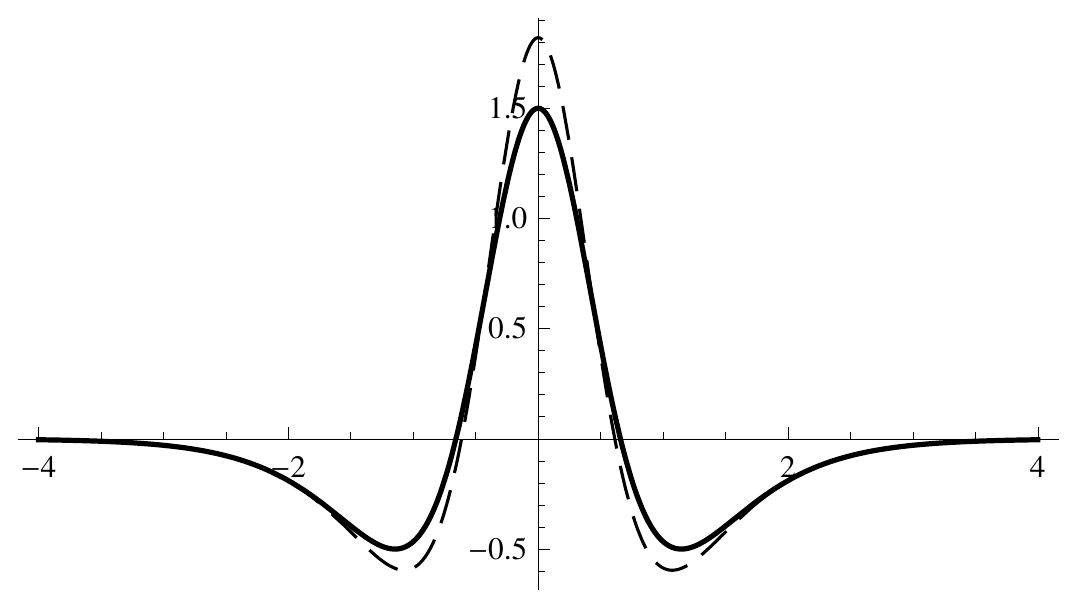}
\end{center}
\caption{\small{The energy density \eqref{eq71}, depicted for $\alpha\!=\!0$ (solid line) and $\alpha\!=\!0.005$ (dashed line).\label{fig14}}}
\end{figure}

One can also express the energy density as follows
\ben\label{eq71}
\rho(y)\!\!\!&=&\!\!\!-(3\!-\!20\alpha)~S^2\!+\!\frac32(3\!-\!232 \alpha)~S^4\!+\!392\alpha~S^6.\;\;\;
\een
Note that in this case the split happens for negative values of $\alpha$, which are not allowed on physical grounds. In Fig.~\ref{fig14} we depict the behavior of the energy density for $\alpha\!=\!0$ (solid line)  and $\alpha\!=\!0.005$ (dashed line).

As stated before, the inclusion of the term $X^2$ does not contribute to destabilize the brane, and the stability behavior is similar to that obtained for the first example in Sec.~III A.

\section{COMMENTS AND CONCLUSIONS}\label{end} 

In this work we have studied thick braneworld models with generalized gravitational dynamics of the form $F=R+\alpha R^n$, and have found analytical solutions for $n=2,3$ and $4$. As explicitly shown, as $n$ increases, the search for analytic solutions becomes more and more complicated. However, we have been able to show that parameterizing the warp factor as the hyperbolic secant of the extra dimension very much simplifies the equations and allows to find analytic solutions in explicit form. This procedure also works when the kinetic term of the scalar field  sourcing the thick brane scenario is changed, which further supports the strength of our method. 

It should be noted that the field equations that determine the form of the scalar field and its potential generically involve up to fourth-order derivatives of the warp function, while our prototypical ansatz \eqref{warpfact} only depends on two free  parameters. This means that there is still freedom to further generalize the warp function ansatz to include two more parameters. Though this possibility might have some interest from a mathematical point of view, we believe that our choice with two parameters provides enough freedom to study the relevant physics of these models, namely, to show that consistent exactly solvable models exist and that they are stable against tensorial perturbations.

The results presented in this work motivate us to suggest that one uses the same methodology to find exact solutions in other scenarios of current interest, with gravity modified as in the Born-Infeld, Gauss-Bonnet and in other cases of current interest.  

DB, LL and RM would like to thank CNPq for partial financial support. The work of GJO is supported by a Ram\'{o}n y Cajal contract, the Spanish grant FIS2011-29813-C02-02, the Consolider Program CPANPHY-1205388,  and the i-LINK0780 grant of the Spanish Research Council (CSIC). The authors also acknowledge funding support of CNPq project No. 301137/2014-5.


\end{document}